\documentclass[english]{article}
\usepackage[affil-it]{authblk}

\usepackage[T1]{fontenc}
\usepackage[latin9]{inputenc}
\usepackage{refstyle}
\usepackage{color,hyperref}
    \catcode`\_=11\relax
    \newcommand\email[1]{\_email #1\q_nil}
    \def\_email#1@#2\q_nil{%
      \href{mailto:#1@#2}{{\emailfont #1\emailampersat #2}}
    }
    \newcommand\emailfont{\sffamily}
    \newcommand\emailampersat{{\color{red}\small@}}
    \catcode`\_=8\relax

\makeatletter

\RS@ifundefined{subref}
  {\def\RSsubtxt{section~}\newref{sub}{name = \RSsubtxt}}
  {}
\RS@ifundefined{thmref}
  {\def\RSthmtxt{theorem~}\newref{thm}{name = \RSthmtxt}}
  
  {}
\RS@ifundefined{lemref}
  {\def\RSlemtxt{lemma~}\newref{lem}{name = \RSlemtxt}}
  {}

\usepackage{amsmath}

\makeatother

\usepackage{babel}
\usepackage{mathrsfs}
\begin{document}
\title{The Schwinger action principle for classical systems }
\author{A. D. Berm\'udez Manjarres}
\affil{\footnotesize Universidad Distrital Francisco Jos\'e de Caldas\\ Cra 7 No. 40B-53, Bogot\'a, Colombia\\ \email{ad.bermudez168@uniandes.edu.co}}
\maketitle
\begin{abstract}
We use the Schwinger action principle to obtain the  equations
of motion in the Koopman-von Neumann operational version of classical
mechanics. We restrict our analysis to non-dissipative systems. We show that for velocity-independent forces the Schwinger action principle can be interpreted as a variational principle.
\end{abstract}

\section{Introduction}

It is well known that classical mechanics has many so-called principles
from where the solution to dynamical problems can be investigated. Among
others, the most famous ones are D'Alembert's principle, Hamilton's
principle, and Gauss's principle of least constraints \cite{var1,var2,var3}.
They all have their advantages and disadvantages from a theoretical
and a practical point of view.

On the other hand, it is perhaps less known that classical mechanics
can be formulated as a theory of operators acting on a Hilbert space.
This idea goes back to Koopman and von Neumann \cite{KvN1,KvN2},
and it is usually referred to as the KvN theory (see \cite{KvN3}
for a review). The KvN theory has been used to study the relations
between quantum and classical mechanics \cite{QC,QC2,QC3,QC4,QC5}.
Moreover, tools originally developed for quantum mechanics have been
applied to purely classical systems \cite{app1,app2,app3,app4,app5,app6}.
The KvN formalism has also received attention in the context of quantum-classical
hybrid theories \cite{hybrid,hybrid2,hybrid3,hybrid4,hybrid5}.

Following the idea of investigating quantum concepts in classical
scenarios, in this paper we explore the Schwinger action principle
\cite{S,S2} in the context of operational classical mechanics.

The Schwinger action principle can be regarded
as the differential form of Feynman path integrals \cite{variation 0},
and it has applications in quantum mechanics and quantum field
theory \cite{variation1,variation2,variation3}.

We will show that the Schwinger action principle can be applied to
classical mechanics and that it leads to the the KvN  equations
of motion. We will also show that, despite being written in terms
of operators, the Schwinger action principle can be interpreted as
a variational principle in the special case of velocity-independent forces. However, the action that is made stationary
is not a functional of paths in the tangent bundle of configuration
space, as is the case for Hamilton's principle. The above makes the
Schwinger action principle unlike any other method from analytical
mechanics.

This paper is organized as follows: in the next section we present
the necessary concepts of the KvN theory. We will follow the standard
treatments where the KvN theory is build up from Hamiltion mechanics
(though we point out that this is not the only possibility, a operational
formulation of classical mechanics can be obtained without the use
of tools from analytical mechanics \cite{QC,AD1,AD2}).

In the last section we define what we understand as a variation of
the dynamical quantities, and we derive the classical dynamical equations
in operator form from the Hamiltonian action as well as the Schwinger
action.

\section{Review of operational classical mechanics}

Let $\Gamma$ stand for the phase space of a classical particle. $\Gamma$
can be endowed with a Hilbert space structure such that time evolution
can be understood as a unitary process. Consider a typical Hamiltonian
of a single particle with a velocity-independent interaction 
\begin{equation}
H=\frac{\mathbf{p}^{2}}{2m}+\phi(\mathbf{r}).
\end{equation}
From classical statistical mechanics, the probability density in phase
space obeys the Liouville equation 

\begin{equation}
\frac{\partial\rho}{\partial t}=-\left\{ \rho,H\right\} ,\label{L}
\end{equation}
where $\left\{ ,\right\} $ stands for the Poisson bracket. The operator
$\left\{ ,H\right\} $ only contains first derivatives. Thus, if a
complex function is defined by $\rho=\left|\psi(\mathbf{r},\mathbf{p})\right|^{2}$,
then $\psi$ will also obey the Liouville equation

\begin{equation}
\frac{\partial\psi}{\partial t}=-\left\{ \psi,H\right\} .\label{scro EQ}
\end{equation}

By its definition, the wavefunction $\psi$ is square integrable in
$\Gamma$. The set of all square integrable complex functions over
phase space defines a Hilbert space $L^{2}(\Gamma)$ where the inner
product is given by integration over the whole phase space

\begin{equation}
\left\langle \varphi,\psi\right\rangle =\int_{\varGamma}\varphi^{*}\psi\,d\mathbf{r}d\mathbf{p}.
\end{equation}
Furthermore, defining the so-called Liouvillian operator

\begin{align}
\hat{L} & =-i\left\{ ,H\right\} ,\label{PB-1}
\end{align}
we can rewrite Eq. (\ref{scro EQ}) as a Schr\"odinger equation
\begin{equation}
\frac{\partial\psi}{\partial t}=i\hat{L}\psi.
\end{equation}

Now, let us define the following differential operators 

\begin{align}
\hat{\lambda}_{\mathbf{r}} & =-i\left\{ ,\mathbf{p}\right\} =-i\nabla_{\mathbf{r}},\nonumber \\
\hat{\lambda}_{\mathbf{p}} & =i\left\{ ,\mathbf{r}\right\} =-i\nabla_{\mathbf{p}}.\label{lambda}
\end{align}
With the help of the operators (\ref{lambda}), we can rewrite the
Liouvillian (\ref{PB-1}) as 

\begin{align}
\hat{L} & =\frac{1}{m}\mathbf{p}\cdot\hat{\lambda}_{\mathbf{r}}-\nabla_{\mathbf{r}}\phi\cdot\hat{\lambda}_{\mathbf{p}}\nonumber \\
 & =\frac{1}{m}\mathbf{p}\cdot\hat{\lambda}_{\mathbf{r}}+\mathbf{F}(\mathbf{r})\cdot\hat{\lambda}_{\mathbf{p}}\label{FULL L}
\end{align}

The above are the basic components of the KvN theory. We can see that
the KvN theory is statistical in nature since $\psi(\mathbf{r},\mathbf{p})$
represents the probability amplitude of finding a particle in a certain
region of phase space. However, particle dynamics is recovered when
point-like delta functions are considered.

The formalism presented so far can be considered as a ``wave mechanics''
version of a more general Dirac bra-ket theory. In its Dirac form,
the position and momentum $(\hat{\mathbf{r}},\hat{\mathbf{p}})$ are
understood as Hermitian operators that commute with each other (but
are not necessarily multiplication operators), and, therefore, there
is no uncertainty between them. The operators $\hat{\lambda}_{\mathbf{r}}$
and $\hat{\lambda}_{\mathbf{p}}$ are defined by its commutation relations
and are understood to have no direct physical meaning (what Sudarshan
calls hidden variables \cite{Sudarshan}, and they are also related
to the Bopp operators in the Wigner-phase space formulation of
quantum mechanics \cite{bopp,bopp2}). It is generally understood that all physically relevant quantities
must be functions only of the complete set of commuting operators $(\hat{\mathbf{r}},\hat{\mathbf{p}})$
and cannot depend on $(\hat{\lambda}_{\mathbf{r}},\hat{\lambda}_{\mathbf{p}})$.
The following commutation relations are satisfied:

\begin{align}
\left[\hat{x}_{i},\hat{x}_{j}\right] & =\left[\hat{p}_{i},\hat{p}_{j}\right]=\left[\hat{x}_{i},\hat{p}_{j}\right]=[\hat{x}_{i},\hat{\lambda}_{p_{j}}]=[\hat{p}_{i},\hat{\lambda}{}_{x_{j}}]=0,\nonumber \\
 & [\hat{\lambda}{}_{x_{j}},\hat{\lambda}_{p_{j}}]=0;\quad[\hat{x}_{i},\hat{\lambda}{}_{x_{j}}]=[\hat{p}_{i},\hat{\lambda}_{p_{j}}]=i\delta_{ij}.\label{comm}
\end{align}
These operators act on a Hilbert space $\mathcal{H}_{cl}$ whose elements
are vectors of the form 
\begin{equation}
\left|\psi\right\rangle =\int\left\langle \mathbf{r},\mathbf{p}\right|\left.\psi\right\rangle \left|\mathbf{r},\mathbf{p}\right\rangle \,d\mathbf{r}d\mathbf{p},
\end{equation}
 such that the complex wavefunction $\psi(\mathbf{r},\mathbf{p})=\left\langle \mathbf{r},\mathbf{p}\right.\left|\psi\right\rangle $
is square integrable. The kets $\left|\mathbf{r},\mathbf{p}\right\rangle $
are eigenfunctions of $\hat{\mathbf{r}}$ and $\hat{\mathbf{p}}$ 

\begin{align}
\hat{x}_{i}\left|\mathbf{r},\mathbf{p}\right\rangle  & =x_{i}\left|\mathbf{r},\mathbf{p}\right\rangle ,\nonumber \\
\hat{p}_{i}\left|\mathbf{r},\mathbf{p}\right\rangle  & =p_{i}\left|\mathbf{r},\mathbf{p}\right\rangle ,
\end{align}
 they obey the orthonnormality condition $\left\langle \mathbf{r}',\mathbf{p}'\right.\left|\mathbf{r},\mathbf{p}\right\rangle =\delta(\mathbf{r}-\mathbf{r}')\delta(\mathbf{p}-\mathbf{p}')$,
and the completeness relation $\int\left|\mathbf{r},\mathbf{p}\right\rangle \left\langle \mathbf{r},\mathbf{p}\right|\,d\mathbf{r}d\mathbf{p}=1$.

In the bra-ket notation, the time evolution equation is written as

\begin{equation}
\frac{d}{dt}\left|\psi(t)\right\rangle =-i\hat{L}\left|\psi(t)\right\rangle ,\label{scrho}
\end{equation}
with 

\begin{equation}
\hat{L}=\frac{1}{m}\hat{\mathbf{p}}\cdot\hat{\lambda}_{\mathbf{r}}+\mathbf{F}(\hat{\mathbf{r}})\cdot\hat{\lambda}_{\mathbf{p}}.\label{L-1}
\end{equation}
Notice that the acceleration 
\begin{equation}
\hat{\mathbf{a}}=i\left[\hat{L},\frac{1}{m}\hat{\mathbf{p}}\right]=\frac{1}{m}\mathbf{F}(\hat{\mathbf{r}})\label{a}
\end{equation}
is a function only of $\hat{\mathbf{r}}$ , and it is then an observable
quantity. 

We can re-express the formalism given so far in terms of a velocity operator instead of the canonical momentum. In cartesian coordinates,
and as we are not yet considering vector potentials, the relation between velocity and momentum is given by

\begin{equation}
\hat{\mathbf{v}}=\frac{1}{m}\hat{\mathbf{p}};\quad\hat{\lambda}_{\mathbf{v}}=m\hat{\lambda}_{\mathbf{p}}.\label{v y p}
\end{equation}
The transformation (\ref{v y p}) is not as ad hoc as it might seem
at first sight, it can be justified from first principles \cite{AD1}.

However, the real advantage of the velocity representation
over the momentum representation comes when we consider velocity-dependent
forces. Using definition (\ref{PB-1}), a Hamiltonian of the form $H=\frac{1}{2m}\left(\mathbf{p}-\mathbf{A}(\mathbf{r})\right)^{2}+\phi(\mathbf{r})$
leads to a very complicated Liouvillian \cite{minimal}. However, the situation greatly
simplifies in the velocity representation. In this case, the transformation
equations are 

\begin{align}
\hat{\mathbf{v}} & =\frac{1}{m}\hat{\mathbf{p}}-\frac{1}{m}{\mathbf{A}}(\hat{\mathbf{r}}),\quad\hat{\lambda}_{\mathbf{v}}=m\hat{\lambda}_{\mathbf{p}},\nonumber \\
\hat{\lambda}_{\mathbf{r}}' & =\hat{\lambda}_{\mathbf{r}}+\frac{\partial}{\partial\hat{\mathbf{r}}}\left({\mathbf{A}}(\hat{\mathbf{r}})\cdot\hat{\lambda}_{\mathbf{p}}\right).\label{canonicalT}
\end{align}
Eqs. (\ref{canonicalT}) define a quantum canonical transformation
as understood in \cite{canonical}, i.e., the transformation preserves the fundamental
bracket relations 

\begin{align}
\left[\hat{x}_{i},\hat{x}_{j}\right] & =\left[\hat{v}_{i},\hat{v}_{j}\right]=\left[\hat{x}_{i},\hat{v}_{j}\right]=[\hat{x}_{i},\hat{\lambda}_{v_{j}}]=[\hat{v}_{i},\hat{\lambda}{}_{x_{j}}']=0,\nonumber \\{}
[\hat{\lambda}{}_{x_{j}},\hat{\lambda}_{v_{j}}] & =0;\quad[\hat{x}_{i},\hat{\lambda}{}_{x_{j}}']=[\hat{v}_{i},\hat{\lambda}_{v_{j}}]=i\delta_{ij}.\label{comm2}
\end{align}
Unitary transformations are canonical, but the converse is not true 
in general \cite{canonical}. The transformation $\left(\hat{\mathbf{r}},\hat{\mathbf{p}},\hat{\lambda}_{\mathbf{r}},\hat{\lambda}_{\mathbf{p}}\right)\rightarrow\left(\hat{\mathbf{r}},\hat{\mathbf{v}},\hat{\lambda}_{\mathbf{r}}',\hat{\lambda}_{\mathbf{v}}\right)$
can be given as a composition of a scale transformation and a unitary
transformation. We refer to Ref \cite{AD1} for more details.

In the velocity representation the Liouvillian takes the form 
\begin{equation}
\hat{L}=\hat{\mathbf{v}}\cdot\hat{\lambda}_{\mathbf{r}}'+\frac{1}{m}\left(\mathbf{F}(\hat{\mathbf{v}},\hat{\mathbf{\mathbf{r}}})\cdot\hat{\lambda}_{\mathbf{v}}\right)_{+},\label{Fxv}
\end{equation}
where 
\begin{equation}
\mathbf{F}(\hat{\mathbf{v}},\hat{\mathbf{\mathbf{r}}})=\left(\frac{\partial}{\partial\hat{\mathbf{r}}}\times\hat{\mathbf{A}}\right)\times\hat{\mathbf{v}}-\frac{\partial\phi(\hat{\mathbf{r}})}{\partial\hat{\mathbf{r}}}
\end{equation}
and we have used the notation $\left(\hat{A}\hat{B}\right)_{+}=\frac{1}{2}\left(\hat{A}\hat{B}+\hat{B}\hat{A}\right)$.

For simplicity, and as we will work solely in the velocity representation,
we will omit the $'$ and we will simply write $\hat{\lambda}_{\mathbf{r}}$
from now on.
In accordance to \cite{AD2}, but unlike \cite{AD1}, we also make
the definition $\left|\mathbf{r},\mathbf{p}\right\rangle \equiv\left|\mathbf{r},\mathbf{v}\right\rangle $. 

Finally, let us mention that the sets $\left\{ \hat{\mathbf{r}},\hat{\mathbf{p}},\hat{\lambda}_{\mathbf{r}},\hat{\lambda}_{\mathbf{p}}\right\} $
and $\left\{ \hat{\mathbf{r}},\hat{\mathbf{v}},\hat{\lambda}_{\mathbf{r}},\hat{\lambda}_{\mathbf{v}}\right\} $
are irreducible in $\mathcal{H}_{cl}$.

\section{Action principles and infinitesimal variations}

From now on, we specialize in systems consisting of a single particle
moving in one dimension since the generalization of the results is immediate.

Just as in quantum mechanics, the dynamics  in the Heisenberg picture
is encoded in the operators via the unitary transformation

\begin{align}
\hat{x}(t) & =U^{\dagger}(t)\hat{x}U(t),\nonumber \\
\hat{v}(t) & =U^{\dagger}(t)\hat{v}U(t),\label{Uevolution}
\end{align}
where the time evolution operator is given by the Schr\"odinger equation
\begin{equation}
i\frac{\partial}{\partial t}U(t)=\hat{L}U(t).
\end{equation}
The evolution of $\hat{\lambda}_{x}(t)$ and $\hat{\lambda}_{v}(t)$
is defined analogously. The evolution of the operators is governed
by Heisenberg equations

\begin{align}
\frac{d\hat{x}}{dt} & =i\left[\hat{L},\hat{x}\right],\quad\frac{d\hat{v}}{dt}=i\left[\hat{L},\hat{v}\right].\label{heisenberg}
\end{align}
Notice that the Liouvillian (\ref{Fxv}) leads to the Newton equations

\begin{equation}
\frac{d\hat{x}}{dt}=\hat{v},\quad\frac{d\hat{v}}{dt}=\frac{1}{m}\hat{F}(\hat{x},\hat{v}).\label{newton Eq}
\end{equation}

We can think of $\hat{x}(t)$ and $\hat{v}(t)$ as defining a path
in operator space, and then we can consider a slightly deformed path
given by
\begin{align}
\hat{x}_{\epsilon}(t) & =\hat{x}(t)+\epsilon\eta(t),\nonumber \\
\hat{v_{\epsilon}}(t) & =\hat{v}(t)+\epsilon\dot{\eta}(t),\label{variation}
\end{align}
for some differentiable function $\eta(t)$ with $\eta(t_{1})=\eta(t_{2})=0$.  Due to the equal time commutation
relations

\begin{equation}
[\hat{x}(t),\hat{\lambda}{}_{x}(t)]=[\hat{v}(t),\hat{\lambda}_{v}(t)]=i,\label{equatl time}
\end{equation}
the deformation (\ref{variation}) can be written as

\begin{equation}
\hat{x}_{\epsilon}(t)=e^{i\epsilon G_{\eta}}\hat{x}(t)e^{-i\epsilon G_{\eta}},\:\hat{v}_{\epsilon}(t)=e^{i\epsilon G_{\eta}}\hat{v}(t)e^{-i\epsilon G_{\eta}}
\end{equation}
where the generator $G_{\eta}$ is given by
\begin{equation}
G_{\eta}=\eta(t)\hat{\lambda}{}_{x}(t)+\dot{\eta}(t)\hat{\lambda}_{v}(t).
\end{equation}
The evolution of $\hat{\lambda}_{x}(t)$ and $\hat{\lambda}_{v}(t)$
is unperturbed under the action of $G_{\eta}$. More general operator variations have been considered \cite{S2},
but we will not consider them here

\subsection{Hamilton principle}

Let us define the Hamiltonian action and the Lagrangian operator by

\begin{equation}
\hat{W}_{H}=\int_{t_{1}}^{t_{2}}\mathscr{\hat{L}}_{H}(\hat{x}(t),\hat{v}(t))dt.
\end{equation}
We consider variations of the form (\ref{variation}) and demand that
$\delta\hat{W}_{H}=0.$ As $\hat{x}(t)$ and $\hat{v}(t)$ commute, we can carry out algebraic manipulations of quantities that only involve these operators just as if they were ordinary functions. Hence,
following similar steps as in calculus of variation, it can be concluded
that a necessary condition for $\delta\hat{W}_{H}=0$ is that $\mathscr{\hat{L}}_{H}$
obeys an equation of the Euler-Lagrange type

\begin{equation}
\frac{d}{dt}\frac{\partial\mathscr{\hat{L}}_{H}}{\partial\hat{v}}-\frac{\partial\mathscr{\hat{L}}_{H}}{\partial\hat{x}}=0.\label{EL}
\end{equation}
For example,  Eq (\ref{EL}) and the Lagrangian
\begin{equation}
\mathscr{\hat{L}}_{H}=\frac{1}{2}m\hat{v}^{2}-\hat{\phi}(\hat{x}),\label{lagrangian}
\end{equation}
 lead directly to (\ref{newton Eq}).
Furthermore, notice that the standard interpretation of the Hamilton's principle is recovered for the eigenvalue equation

\begin{equation}
\mathscr{\hat{L}}_{H}(\hat{x}(t),\hat{v}(t))\left|x,v\right\rangle =\mathscr{L}_{H}(x(t),v(t))\left|x,v\right\rangle .
\end{equation}
The physical trajectory of the particle is the one that makes
stationary the action integral $\int_{t_{1}}^{t_{2}}\left[\left\langle x',v'\right|\mathscr{\hat{L}}_{H}(\hat{x}(t),\hat{v}(t))\left|x,v\right\rangle \right]dt$.

\subsection{Schwinger action principle}

While the equations of the last subsection are written in terms of
operators, it is true that they just mimic the standard equations
of Lagrangian mechanics and ultimately do not give any new insight.
The procedure we will give now is of a different nature, and, to the
author's best knowledge, represents a new action principle for classical
systems.

Let $\left|1\right\rangle$  and $\left|2\right\rangle$  be two states defined by the values of a complete set of commuting operators at two different times. Notice that for the single particle $(\hat{x},\hat{v})$, $(\hat{x},\hat{\lambda}_{v})$, $(\hat{\lambda}_{x},\hat{v})$ and $(\hat{\lambda}_{x},\hat{\lambda}_{v})$ are valid choices for the complete set of commuting operators. Following Schwinger, a slight variation on the operators, such as the one given by (\ref{variation}),  produces a variation in their eigenstates such that
\begin{equation}
\delta\left\langle 1\right|\left.2\right\rangle =\left\langle 1\right|\delta\hat{W}_{S}\left|2\right\rangle,\label{Saction0} 
\end{equation}
where the action operator  is given by
\begin{equation}
\hat{W}_{S}=\int_{t_{1}}^{t_{2}}\left(\left(\hat{\lambda}_{x}\frac{d\hat{x}}{dt}+\hat{\lambda}_{v}\frac{d\hat{v}}{dt}\right)_{+}-\hat{L}\right)dt,\label{Saction}
\end{equation}
The symmetrization in (\ref{Saction}) ensures Hermiticity
in the general case of velocity-dependent forces. 

A variation of the endpoints has information about the canonical commutation
relations between the operators of the theory but we will maintain
the endpoint fixed since we are taking Eqs (\ref{comm}) and  (\ref{comm2}) for granted.
The variation with fixed endpoints (\ref{variation}) of the action
operator (\ref{Saction}), this is, a variation such that $\delta\hat{W}_{S}=0$,
 gives the following Hamilton-like equation of motion \cite{S,S2,variation1,variation2,variation3} 

\begin{align}
\frac{d\hat{x}}{dt} & =\frac{\partial\hat{L}}{\partial\hat{\lambda}_{x}},\quad\frac{d\hat{v}}{dt}=\frac{\partial\hat{L}}{\partial\hat{\lambda}_{v}},\label{hamiltonlike}\\
\frac{d\hat{\lambda}_{x}}{dt} & =-\frac{\partial\hat{L}}{\partial\hat{x}},\quad\frac{d\hat{\lambda}_{v}}{dt}=-\frac{\partial\hat{L}}{\partial\hat{v}},\\
\frac{d\hat{L}}{dt} & =\frac{\partial\hat{L}}{\partial t}.
\end{align}
With the help of commutation relations (\ref{equatl time}) and the Liouvillian (\ref{Fxv}), the  Eqs (\ref{hamiltonlike})
can be rewritten as the Heisenberg equations 
\begin{align}\frac{d\hat{x}}{dt}=\hat{v},\quad\frac{d\hat{v}}{dt}=\frac{1}{m}\hat{F}(\hat{x},\hat{v}).
\end{align}
Hence, the KvN formalism and the whole classical dynamics are
recovered.

Replacing the Liouvillian (\ref{Fxv}) into (\ref{Saction}), we can rewrite the Schwinger
action as

\begin{equation}
\hat{W}_{S}=\int_{t_{1}}^{t_{2}}\mathscr{\hat{L}}_{S}dt
\end{equation}
with the Schwinger ``Lagrangian'' operator given by

\begin{equation}
\mathscr{\hat{L}}_{S}=\left(\left(\frac{d\hat{v}}{dt}-\frac{1}{m}F(\hat{x},\hat{v})\right)\hat{\lambda}_{v}\right)_{+}.
\end{equation}

The above can easily be generalized to the case of several particles. Using an upper
index to identify different particles, $\mathscr{\hat{L}}_{S}$ can
 be written  as

\begin{equation}
\mathscr{\hat{L}}_{S}=\sum_{i}\left(\left(\frac{d\hat{\mathbf{v}}^{(i)}}{dt}-\frac{1}{m_{i}}\mathbf{F}^{(i)}(\hat{\mathbf{r}},\hat{\mathbf{v}})\right)\cdot\hat{\lambda}_{\mathbf{v}}^{(i)}\right)_{+}.
\end{equation}

Finally, the Schwinger action principle has an interpretation as a stationary principle for the special case of velocity-independent forces, though in a more abstract sense than Hamilton's. Consider the complete set of orthonormal
kets $\left|x,\lambda_{v}\right\rangle $ obtained from $\left|x,v\right\rangle $
by the relation

\begin{equation}
\left\langle x',v'\right.\left|x,\lambda_{v}\right\rangle =\frac{1}{\sqrt{2\pi}}\delta(x-x')\exp[v'\lambda_{v}].
\end{equation}
The $(x,\lambda_{v})$ representation was considered in \cite{gozzi},
see also \cite{KvN3}. The kets $\left|x,\lambda_{v}\right\rangle $
are eigenvectors of the complete set of commuting operators ($\hat{x}$,$\hat{\lambda}_{v}$),
hence, for velocity-independent forces, we have 

\[
\mathscr{\hat{L}}_{S}(\hat{x}(t),\hat{\lambda}_{v}(t))\left|x,\lambda_{v}\right\rangle =\mathscr{L}_{S}(x(t),\lambda_{v}(t))\left|x,\lambda_{v}\right\rangle .
\]
The physical acceleration $\frac{d\hat{v}}{dt}$ is the one that
makes stationary the action integral $\int_{t_{1}}^{t_{2}}\left[\left\langle x',\lambda_{v}'\right|\mathscr{\hat{L}}_{S}(\hat{x}(t),\hat{\lambda}_{v}(t))\left|x,\lambda_{v}\right\rangle \right]dt$.

\section{Conclusions}

We have shown that the equations of motion of classical mechanics are derivable from the Schwinger action principle. An interpretation as a stationary principle in the $(x,\lambda_{v})$ space is possible for velocity-independent forces. The interpretation for forces that do
depend on $\hat{\mathbf{v}}$ remains unclear since $\hat{\mathbf{v}}$
and $\hat{\lambda}_{\mathbf{v}}$ are incompatible operators in the
quantum sense. It is worth pointing out that the operators $(\hat{\lambda}_{x},\hat{\lambda}_{v})$, usually taken as hidden variables, play a central role in the results given in the paper; this is in sharp contrast to Hamilton's principle where only observable operators appear. Another crucial difference is that the Schwinger ``Lagrangian'' operator is given directly in terms of the forces and not in terms of the potentials. The relation between Schwinger's and Hamilton's principles requires further investigation.

\end{document}